\documentclass[11pt]{article}
\usepackage[preprint]{acl}
\usepackage{times}
\usepackage{latexsym}
\usepackage[T1]{fontenc}
\usepackage[utf8]{inputenc}
\usepackage{microtype}
\usepackage{inconsolata}
\usepackage{graphicx}
\usepackage{amsmath}
\usepackage{amssymb}
\usepackage{booktabs}
\usepackage{tabularx}
\usepackage{array}
\usepackage{url}
\usepackage{placeins}

\newif\ifnamedversion
\namedversiontrue
\newif\ifappendixnow

\title{Calibration-Induced Degeneracy in LLM Financial Forecasting:\\
An Audit-Trailed Case Study on Next-Day Market Risk}

\author{Arin Mohanty\\
  Stony Brook University\\
  Stony Brook, New York, USA}

\begin{document}
\maketitle

\begin{abstract}
Costly LLM features are useful only if calibration leaves them a path into the
forecast. In a next-day risk experiment for two broad-market exchange-traded
funds, full-history scoring preceded the 2022 calibration. All four LLM
importance weights reached zero; the 856 later scores therefore could not
affect the evaluation. We call this \emph{calibration-induced degeneracy}.

A signed repair activated all four mappings but produced no improvement after
familywise correction. A near-zero-cost headline count, however, improved SPY
continuous variance by 0.001720, with a 95\% familywise interval of
$[0.000719,0.002830]$. The contrast shows that an expensive semantic feature
must earn its place against a trivial control before full-history acquisition.

We propose a calibration-viability checkpoint: fit the mapping, perturb the
feature over prespecified calibration values, and acquire holdout values only
if the forecast changes by a meaningful threshold. The check uses no holdout
outcomes and would have stopped this study before its \$16.17 full-history
inference phase.

\end{abstract}

\appendixnowfalse
\ifappendixnow\else

\section{Introduction}\label{sec:introduction}

LLMs increasingly generate bespoke features for financial forecasting, but
feature acquisition is justified only if the calibrated model can use them. A
standard protocol moves from prompt development to calibration, paid
full-history scoring, and out-of-sample evaluation. One inexpensive question
must separate calibration from acquisition: \emph{does the fitted forecast
still respond to the feature?} Its answer determines whether the experiment
should proceed to paid scoring.

This paper documents that failure in an internally prespecified, hash-frozen
experiment. An LLM read daily headlines and assigned an importance score for
broad U.S. stocks. The planned test asked whether that score improved next-day
SPY risk forecasts beyond an option-implied-volatility baseline; QQQ/VXN
supplied a correlated transfer check. The project acquired 856 usable later
scores before executing the 2022 calibration, which assigned all four news
weights zero. We call the resulting sequencing failure
\emph{calibration-induced degeneracy}.

The usual safeguards did not catch the failure. The mapping used only
calibration outcomes, later outcomes remained out of sample for its
coefficients, and zero was a legitimate nested-model choice. Those rules
protect against look-ahead and overfitting; they do not guarantee that the
feature remains active.

A separate limitation concerns prior exposure. Earlier project work had shown
the author SPY outcomes from the later calendar period, creating an
experimenter-expectancy risk outside the coefficient fit. QQQ/VXN outcomes
remained unopened before v4 scoring and provide the cleaner untouched control,
although not an independent replication. Section~\ref{sec:data} gives the full
timing record.

The diagnostic contrast deliberately starts with the simplest plausible
feature: headline volume. A signed repair restored an active LLM mapping but
produced no improvement after correction across four endpoints. A standardized
headline count---a near-zero-cost feature---improved SPY continuous variance by
0.001720, with a 95\% familywise interval of $[0.000719,0.002830]$. The count
was mixed elsewhere, but its role is decisive: an expensive semantic feature
must earn its place against a trivial volume measure.

The solution is a mandatory \emph{calibration-viability checkpoint}: fit the
frozen mapping, perturb the candidate feature over prespecified values, and
acquire holdout features only after a meaningful forecast response. The check
uses no holdout outcomes. Here it would have stopped the \$16.17 full-history
inference phase. That amount is an operational illustration, not an economic-
loss estimate; the avoidable cost scales with the breadth of the scoring
program.

Detailed prompt, retry, correction, and verification records remain in the
supplement. The design was internally hash-frozen rather than publicly
registered; the diagnostics are post-hoc; exact dates leave memorization
unresolved; and verification was same-author rather than external replication.

\section{Relation to Prior Work}\label{sec:related}

\subsection{Financial Text and LLM Forecasts}

Financial-news research has long asked whether text predicts anything beyond
what prices already reveal. LLM studies extend this question by using either
learned text representations or direct model judgments.
Early text-and-markets work quantified newspaper pessimism and related it to
returns and trading volume \citep{tetlock2007content}. A later long-history
measure learned news-implied volatility from newspaper text and linked it to
disaster concerns and risk premia \citep{manela2017news}. These studies show
that text can measure economically relevant states; they do not imply that one
retrospective daily LLM score must add to a same-day
option-implied-volatility benchmark.
\citet{lopezlira2023chatgpt} report return predictability from LLM headline
interpretation, while \citet{guo2024newsflow} compare two types of learned
news representation for stock-return prediction. \citet{yu2023volatility}
shows that repeated LLM classifications can vary and that this variation can
affect later financial decisions. \citet{wang2025consistency} examines the
same problem across finance and accounting tasks and warns against reporting
only favorable generations. Those studies focus on model outputs; we focus on
the path from a frozen score to a benchmarked forecast.

Three closer lines of work clarify what our negative result does and does not
challenge. Newspaper-based equity-market-volatility trackers can closely
follow realized volatility and VIX \citep{baker2019policy}. Human-coded
next-day newspaper accounts can identify the causes of major international
stock-market jumps \citep{baker2021triggers}. Those are measurement or
explanation exercises, but they do not ask whether one daily LLM score adds to
a fixed next-day VIX forecast. At higher frequency, \citet{he2026risk}
combines contemporaneous news narratives with market data to study systematic
macro risk and reports strong out-of-sample portfolio performance. That
positive design uses different timing, targets, representations, and economic
objects.

We evaluate one overall daily judgment about broad-market importance and
next-day risk, rather than the direction of an individual stock. We compare it
with option-implied volatility and GARCH and trace the entire path from LLM
score to fitted forecast.

\subsection{Forecast Evaluation and Benchmark Choice}

GARCH is a standard benchmark that forecasts future volatility from patterns
in recent returns \citep{bollerslev1986garch}. VIX and VXN instead summarize
volatility implied by option prices. \citet{hansen2005forecast} and
\citet{diebold1995predictive} explain why conclusions depend on both the
benchmark and the scoring rule. We compare forecast losses date by date and
use a moving-block bootstrap, which resamples consecutive 20-day stretches so
that nearby market days are not treated as independent. A later exploratory
check used Student-t GARCH(1,1), a version designed to accommodate
unusually large returns.

Because our candidate forecasts nest their no-news baselines, standard
equal-accuracy inference can behave nonstandardly in such comparisons.
Established work provides specialized tests and loss adjustments for estimated
nested forecasts
\citep{clarkmccracken2001nested,clarkwest2007nested}. Our post-hoc block
intervals instead condition on the fixed 2022 mapping; they are not presented
as Clark--West or Clark--McCracken tests.

\subsection{Prespecification, Specification Search, and Boundaries}

Preregistration records the planned test in a time-stamped, read-only form
before its answer is known, helping readers distinguish confirmation from
later exploration \citep{nosek2018preregistration}. Internal hashes alone
establish file identity, not when a plan existed. Recent capital-markets work
uses externally anchored forward records \citep{zheng2026forward}, while
another audit study explicitly distinguishes a prespecified plan from
timestamp-enforced preregistration \citep{khan2026audit}. We therefore describe
v4 as internally prespecified and hash-frozen, not preregistered.

This distinction matters especially in finance, where testing many
alternatives can produce chance findings \citep{harvey2016crosssection}.
Estimates at a constraint boundary, such as our required nonnegative weight
reaching zero, also need special care \citep{self1987boundary}.
Machine-learning venues have separately tested results-blind plans and
reproducibility checklists
\citep{bertinetto2021preregistration,pineau2021reproducibility}.

Here the boundary estimate created an operational, not merely inferential,
failure inside a cost-capped LLM experiment. Section~\ref{sec:degeneracy}
isolates the mechanism and turns it into a prospective design requirement.

The requirement is closely related to prespecified futility monitoring, which
stops an experiment when continuation cannot answer its question
\citep{lachin2005futility}. Its application to costly generated features is
direct: establish that the calibrated model has a mechanical path from the
feature to the forecast before purchasing the feature at scale.

\section{Calibration Viability as a Design Requirement}\label{sec:viability}

Let $\widehat f_A(x,s)$ denote a calibrated augmented forecast using ordinary
inputs $x$ and candidate feature $s$, and let $\widehat f_B(x)$ denote its
baseline. Nested calibration makes the augmented model degenerate whenever
$\widehat f_A(x,s)=\widehat f_B(x)$ for every plausible $s$. In that state,
holdout feature values are algebraically irrelevant: no sample size, bootstrap
scheme, or protected outcome window can identify their contribution through
the frozen mapping.

Viability must therefore be defined before calibration. For prespecified
anchor inputs $\mathcal X$, a feature-support grid $\mathcal S$, and a
meaningful-change threshold $\delta$, compute
\begin{equation}
D=\max_{x\in\mathcal X}\max_{s,s'\in\mathcal S}
  \left|\widehat f_A(x,s)-\widehat f_A(x,s')\right|.
\label{eq:viability}
\end{equation}
In practice, set $\mathcal S$ to the standardized feature's 10th, 50th, and
90th calibration-sample percentiles, with the observed minimum and maximum as
additional endpoint checks for a linear mapping. Choose $\mathcal X$ from
calibration rows spanning low, median, and high baseline-risk forecasts.
The mapping is mechanically nonviable when $D=0$ and operationally too weak
when $0<D<\delta$. The threshold is not estimated from holdout outcomes; it is
fixed from the forecast scale, scoring rule, and minimum change worth paying
to evaluate.

This requirement changes the acquisition order. Researchers first freeze the
candidate and cheap comparison features, calibration sample, support grid,
$\delta$, and fallback. They then fit and perturb the mappings. Only a mapping
with $D\geq\delta$ proceeds to paid holdout-feature acquisition. Section
\ref{sec:gate} converts this principle into a five-step checkpoint; the next
section shows exactly where the present study omitted it.

\section{Data, Timing, and Autonomous Score}\label{sec:data}

\subsection{News and Market Timing}

The historical news collection spans 2017 through June 2026. Each trading day
contains at most 25 selected headlines, with a cap preventing one outlet from
dominating the packet. Both the 2022 fitting period and the 2023--2026 later
period use the same GDELT source regime \citep{gdelt2014gkg}. Daily SPY, QQQ,
VIX, and VXN data came from the Yahoo Finance chart endpoint. Before evaluation,
we saved the download addresses, raw responses, cleaned files, and digital
fingerprints.

Construction was deterministic. We retained titles from a 26-domain news
whitelist when they matched at least one of 38 fixed U.S.-macro or market
phrases. Within each market-close bucket, titles were normalized by removing
wire-service update prefixes, trailing outlet names, punctuation, and repeated
spaces. We treated two titles as duplicates if they were exact normalized
matches, containment matches of at least 25 characters, or had
\texttt{SequenceMatcher} similarity of at least 0.85. The earliest crawl
observation was kept. Remaining titles were ranked first by the number of
distinct matched phrases, then by crawl time and URL. We selected at most 25,
with no more than five from one base domain. Appendix~\ref{app:news-construction}
lists the exact domains and phrases.

The timestamp is GDELT's crawl-observation time, not a guaranteed publisher
time. Buckets run from the previous NYSE close to the current close in
\texttt{America/New\_York}. The ordinary close is 16:00; prespecified early
closes use 13:00, with daylight-saving conversion to UTC. The v4 years used the
GKG era. No separate post-download language classifier was applied beyond the
English-language outlet/domain collection. Updated stories were not tracked as
evolving article objects. The rule above removed a near-duplicate update, while
a materially different title could remain. These are measurement choices, not
neutral preprocessing.

Each news packet covers the period after the previous market close through the
current close. Its score forecasts the following trading day's return. The
LLM sees the close time, headlines, source domains, observation times, and
anonymous positions. It does not see realized returns, VIX/VXN, outcome
labels, earlier scores, or information about which evaluation group a date
belongs to.

Market snapshots for SPY, QQQ, VIX, and VXN were frozen only after all
2023-2026 LLM outputs were final. VIX or VXN at the origin close precedes the
next-day target. Asset returns use Yahoo adjusted closes,
$r_{t+1}=100(\mathrm{AdjClose}_{t+1}/\mathrm{AdjClose}_t-1)$. Returns are in
percentage points and reflect splits and cash distributions to the extent
encoded in Yahoo's adjusted-close series. Sessions are keyed by the U.S.
trading date; VIX/VXN use the unadjusted index close on the origin date.

\subsection{Autonomous P1 Judgment}

The selected prompt asks what is genuinely important for broad U.S. stocks
over the next one to five trading days. The LLM scores each headline, then
forms one overall daily importance judgment from 0 to 100. It must reason
about the packet as a whole rather than mechanically averaging item scores.
The prompt also told it to distinguish new causes from headlines that merely
describe a market move and not to use later developments or an event's later
historical fame.

We used the scorer in July 2026 with the API label
\texttt{gpt-5.6-luna} and the provider's ``high'' reasoning-effort setting.
We report both verbatim rather than infer equivalence to another public model
family or release vintage. The OpenAI Batch Responses API was used, storage was
false, and temperature was omitted.

The score itself did not collapse to a few values. In 2022 it ranged from 4 to
98 (median 73, mean 70.50, population SD 17.43; 60 distinct values). In the
later sample it ranged from 5 to 97 (median 73, mean 69.41, SD 17.34; 76
distinct values). Full quantiles appear in Appendix~\ref{app:reviewer}.

This measurement still has a major unresolved threat: the LLM sees exact past
dates and may remember famous events. Telling it not to use later historical
fame cannot erase pretraining. Any positive exploratory association could
therefore reflect contemporaneous judgment, memorized outcomes, or both. We
did not run a date-masked re-scoring experiment because that would require a
new prompt and paid outputs.

\subsection{Prior Exposure and Execution Order}\label{sec:chronology}

A critical limitation is prior outcome exposure. A separate Hybrid v1.10
project had already examined realized SPY outcomes through June 2026. The v4
mapping was never fitted on those later outcomes, but the author knew results
from the same SPY calendar period before v4 was complete. That knowledge
creates a nonstandard experimenter-expectancy risk in prompt selection and the
decision to proceed, even though it cannot enter the frozen coefficient fit
mechanically. QQQ/VXN outcomes remained unopened before v4 full-history
scoring and therefore provide the cleaner untouched control. They share dates
and broad shocks with SPY, however, so they are a transfer check rather than an
independent replication.

\begin{figure*}[t]
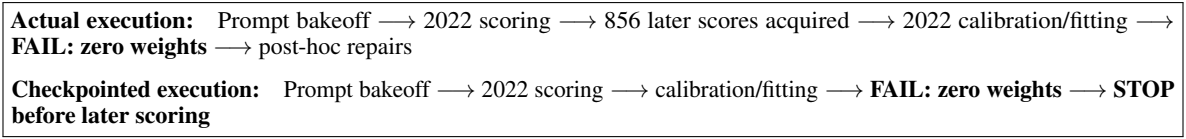

\centering
\fbox{\begin{minipage}{0.96\textwidth}
\small
\textbf{Actual execution:}\quad
Prompt bakeoff $\longrightarrow$ 2022 scoring $\longrightarrow$
856 later scores acquired $\longrightarrow$ 2022 calibration/fitting
$\longrightarrow$ \textbf{FAIL: zero weights} $\longrightarrow$ post-hoc repairs

\medskip
\textbf{Checkpointed execution:}\quad
Prompt bakeoff $\longrightarrow$ 2022 scoring $\longrightarrow$
calibration/fitting $\longrightarrow$ \textbf{FAIL: zero weights}
$\longrightarrow$ \textbf{STOP before later scoring}
\end{minipage}}
\caption{Execution timeline. The actual protocol acquired the 856 later scores
before calibration revealed degeneracy. The checkpoint reorders the stages and
would have stopped before the paid full-history phase.}
\label{fig:timeline}
\end{figure*}

The project used SHA-256 fingerprints to show that plans and files did not
change after freezing. The hashes establish file identity, not an independent
timestamp; accordingly, the v4 design is internally prespecified and
hash-frozen rather than preregistered.

Prompt development used four alternatives on 27 dates without opening market
outcomes. A coverage correction made after inspecting LLM outputs changed the
recorded decision from STOP to GO; it is therefore disclosed as a substantive
procedural correction, not invisible preprocessing.

P1 then scored 2022. The first pass failed its coverage gate. After the market
result was known, exactly the 16 token-truncated outputs were retried under an
already permitted mechanical rule. Final coverage was 250/251 and the rank
correlation was 0.0980 with a 90\% interval $[0.0045,0.1867]$, so the frozen
screen passed.

Before opening any 2023--2026 v4 LLM output or QQQ/VXN outcome, the project
fixed the 865 dates, endpoints, 2022-only fitting rule, retries, bootstrap,
classifications, code hashes, and spending cap. Full-history inference then
produced 856 strict-valid later scores. Only after those scores existed did the
project open the training-only VIX/VXN snapshots and execute the 2022
calibration. That fit assigned all four importance weights zero.

Figure~\ref{fig:timeline} isolates the sequencing failure. Separate
same-author reimplementations reconstructed the main and later analyses; they
are software checks, not external replication. The complete correction,
retry, cost, and verification record is preserved in
Appendix~\ref{app:chronology}.

\section{Prespecified Forecast Design}\label{sec:design}

For each asset, we forecast two next-day endpoints: whether the absolute return
exceeds the trailing five-year 90th percentile, and the squared percent return.
The daily importance score is rescaled as
$z_t=(\mathrm{importance}_t-50)/25$. VIX for SPY and VXN for QQQ provide the
option-implied-volatility baselines.

The news models add one score coefficient to each baseline. Both coefficients
were restricted to $[0,5]$: zero allowed the model to ignore news, while the
positive range encoded the directional hypothesis that more important news
should not reduce next-day risk. Binary models were fit by log loss.
Continuous-variance models used Gaussian quasi-likelihood (QLIKE), which is
robust to noisy variance proxies such as squared returns
\citep{patton2011volatility}. Both models used the 250 usable 2022 dates.
Appendix~\ref{app:models} gives the exact transformations, equations, and
numerical safeguards.

All fitted parameters were then locked and applied without refitting to
2023--2026. Improvement is baseline loss minus news-model loss, so positive
values favor news. We report 95\% intervals from 20,000 moving-block bootstrap
draws using 20-day blocks; 5-, 10-, 40-, and 60-day checks did not alter the
reported classifications. SPY/VIX is primary. QQQ/VXN is a correlated
same-date transfer check, not independent replication.

\section{Main and Contrasting Results}\label{sec:registered-results}

\subsection{Inference Coverage}

\begin{table}[t]
\centering
\small
\caption{Full-history inference coverage. Of the 50 first-pass failures, 44 met
the ceiling-only retry rule; the other six were non-retriable.}
\label{tab:coverage}
\begin{tabular}{@{}lr@{}}
\toprule
Quantity & Result \\
\midrule
Eligible later dates & 865 \\
Passed format rules initially & 815 \\
Ceiling retry candidates & 44 \\
Recovered by retry & 41 \\
Final usable scores & 856 \\
Final invalid & 9 \\
Final coverage & 98.96\% \\
\bottomrule
\end{tabular}
\end{table}

The nine invalid dates were left missing. We did not rewrite malformed
responses, fill in scores, make a second retry, switch prompts, or pay for a
post-result rescue. They are listed and stress-tested in
Appendix~\ref{app:reviewer}.

\subsection{Calibration and Forecast Loss}

Table~\ref{tab:main-contrasts} places the prespecified result beside the two
diagnostics that identify its cause and practical consequence. Panel A reports
the frozen comparison, Panel B activates the LLM score by relaxing only its
sign restriction, and Panel C substitutes the cheap headline-count feature.

\begin{table*}[t]
\centering
\scriptsize
\caption{Main and contrasting results on 856 later dates per asset. Improvement
equals loss under the option-implied-volatility baseline minus feature-model
loss; positive values favor the feature model. Panel A is prespecified. Panels
B and C are post-hoc; the final column gives 95\% Bonferroni familywise
intervals across four comparisons.}
\label{tab:main-contrasts}
\begin{tabular}{@{}lllrrr@{}}
\toprule
Feature/mapping & Asset & Endpoint & 2022 slope & Improvement [95\% interval] & Familywise interval \\
\midrule
\multicolumn{6}{@{}l}{\textit{Panel A: Prespecified nonnegative LLM importance}} \\
LLM, $[0,5]$ & SPY/VIX & Binary & $0$ & $0\ [0,0]$ & --- \\
LLM, $[0,5]$ & SPY/VIX & Variance & $0$ & $0\ [0,0]$ & --- \\
LLM, $[0,5]$ & QQQ/VXN & Binary & $0$ & $0\ [0,0]$ & --- \\
LLM, $[0,5]$ & QQQ/VXN & Variance & $0$ & $0\ [0,0]$ & --- \\
\addlinespace
\multicolumn{6}{@{}l}{\textit{Panel B: Signed LLM repair}} \\
LLM, $[-5,5]$ & SPY/VIX & Binary & $-0.00536$ & $-0.0000211\ [-0.0000721,0.0000313]$ & $[-0.0000855,0.0000473]$ \\
LLM, $[-5,5]$ & SPY/VIX & Variance & $-0.15878$ & $-0.007452\ [-0.015888,-0.001066]$ & $[-0.018435,0.000266]$ \\
LLM, $[-5,5]$ & QQQ/VXN & Binary & $-0.09386$ & $-0.000753\ [-0.001706,0.000258]$ & $[-0.001986,0.000565]$ \\
LLM, $[-5,5]$ & QQQ/VXN & Variance & $-0.17830$ & $-0.008548\ [-0.018895,-0.000751]$ & $[-0.022584,0.000948]$ \\
\addlinespace
\multicolumn{6}{@{}l}{\textit{Panel C: Signed cheap headline count}} \\
Log count & SPY/VIX & Binary & --- & $-0.007913\ [-0.011833,-0.004358]$ & $[-0.013017,-0.003549]$ \\
Log count & SPY/VIX & Variance & --- & $0.001720\ [0.000935,0.002580]$ & $[0.000719,0.002830]$ \\
Log count & QQQ/VXN & Binary & --- & $0.000150\ [0.000011,0.000291]$ & $[-0.000020,0.000332]$ \\
Log count & QQQ/VXN & Variance & --- & $-0.007883\ [-0.012927,-0.003350]$ & $[-0.014661,-0.002344]$ \\
\bottomrule
\end{tabular}
\end{table*}

Panel A reports the prespecified result: all four 2022 importance slopes were
zero, so neither the SPY claim nor the QQQ transfer check was confirmed.
Excluding the twelve prompt-development dates left 844 evaluation dates and
changed nothing.

Panels B and C do not revive the original claim; they diagnose it. The active
signed LLM mapping produced no familywise-corrected gain, while the cheap count
produced one. Sections~\ref{sec:degeneracy} and \ref{sec:exploratory} explain
why that contrast matters.

\section{Calibration-Induced Degeneracy}\label{sec:degeneracy}

\subsection{The Failure Occurred at Calibration}

The decisive event was the 2022 boundary solution, not a noisy holdout
estimate. With every importance coefficient at zero, each augmented forecast
equaled its option-implied-volatility baseline for every input; changing the
score from 0 to 100 changed nothing. Every per-date loss contrast and bootstrap
resample was zero. The 856 later scores therefore could not affect the
evaluation. Appendix~\ref{app:models} gives the equations, and Panel A of
Table~\ref{tab:main-contrasts} gives the empirical result.

\subsection{The Signed Repair Identifies the Boundary Mechanism}

The sign restriction caused the collapse. A post-hoc diagnostic changed only
the importance-slope bounds from $[0,5]$ to $[-5,5]$. It then refitted the same
250 calibration rows and transported those fits without later-period
refitting. Every optimum moved to an interior negative value, as Panel B of
Table~\ref{tab:main-contrasts} shows.

The negative slopes are economically plausible conditional adjustments, not
evidence that important news is intrinsically calming. The prompt asks about
importance over one to five trading days. The target instead measures risk on
the next day, after news has already been collected through the close.
Information can be incorporated before the close, resolve uncertainty, or
unfold beyond one day. The nonnegative restriction encoded a directional
hypothesis that the measurement horizon did not guarantee.

The repair restored the feature path, but all four later-period improvement
estimates were nonpositive. The binary intervals included zero. Both variance
intervals excluded zero in the harmful direction at ordinary 95\% coverage,
but not after correction across four endpoints. The SPY binary training gain
was only $1.12\times10^{-6}$. No endpoint produced a familywise-corrected
improvement.

\subsection{A Valid Model Choice Did Not Create a Valid Feature Test}

Selecting a no-news model on calibration data is a legitimate forecasting
decision. It did not, however, test the acquired feature. The supported
conclusion is correspondingly narrow: the prespecified procedure did not
evaluate incremental score value, not that every possible mapping is
uninformative.

The protocol required at least 95\% valid LLM outputs but never required an
active calibrated mapping. Full-history inference also preceded the calibration
fit. The project therefore discovered the failed viability gate only after
acquiring the costly feature. Section~\ref{sec:gate} supplies the missing stop
rule; exact cost and procedural records remain in
Appendix~\ref{app:chronology}.

\section{The Cheap Baseline as a Critical Diagnostic}\label{sec:exploratory}

The most informative post-hoc contrast is also the least expensive feature.
We replaced LLM importance with the standardized log count of relevant
headlines before deduplication and the top-25 cap, then fitted the same signed
mappings over the option-implied-volatility baselines. The rules were frozen
before calculation, and the analysis required no new LLM calls.

Panel C of Table~\ref{tab:main-contrasts} reports the result. Headline count
improved SPY continuous variance by 0.001720; its 95\% familywise interval,
$[0.000719,0.002830]$, excludes zero. The count harmed SPY binary loss and QQQ
variance and did not produce a familywise-resolved QQQ binary gain. It is
therefore not a generally superior forecast. Its diagnostic force is sharper:
a nearly free packet-level statistic succeeded on one endpoint where the LLM
representation failed both the prespecified gate and the active signed repair.

Raw association does not overturn that contrast. Later-period Spearman
correlations with next-day absolute return were 0.1413 for SPY and 0.1284 for
QQQ, but both fell to about 0.05 after conditioning on VIX/VXN. The diagnostic
question is incremental calibrated value beyond option prices and headline
volume, not unadjusted correlation.

The complete direct-association table, conditional-rank analysis, GARCH grid,
standalone forecast, missing-date bounds, and 16-endpoint sensitivity are
reported once in Appendix~\ref{app:additional}; none changes the central
diagnosis.

\section{A Practical Calibration-Viability Checkpoint}\label{sec:gate}

The operational rule is \textbf{fit first, perturb second, acquire last}.
Figure~\ref{fig:checkpoint} gives the complete decision rule.

\begin{figure*}[t]
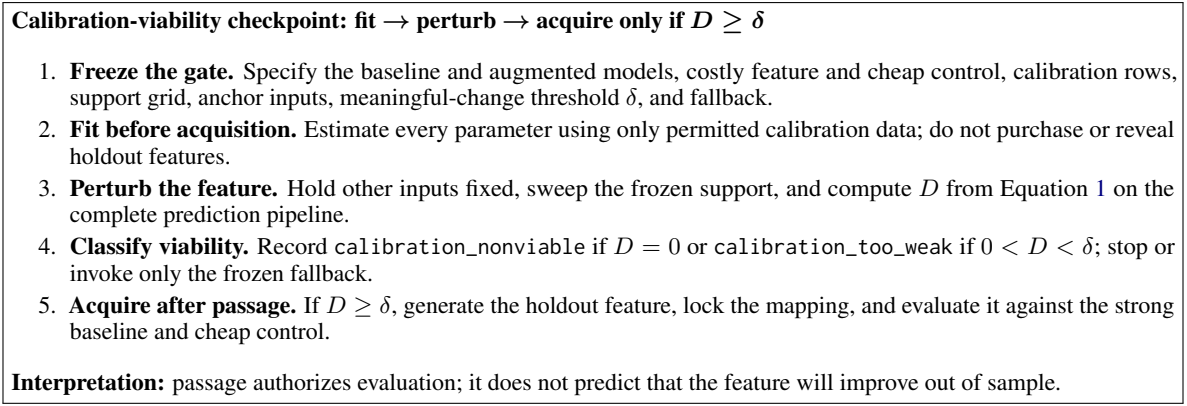

\centering
\fbox{\begin{minipage}{0.96\textwidth}
\small
\textbf{Calibration-viability checkpoint: fit $\boldsymbol{\rightarrow}$
perturb $\boldsymbol{\rightarrow}$ acquire only if $\boldsymbol{D\geq\delta}$}
\begin{enumerate}
\setlength{\itemsep}{2pt}
\setlength{\parskip}{0pt}
\item \textbf{Freeze the gate.} Specify the baseline and augmented models,
  costly feature and cheap control, calibration rows, support grid, anchor
  inputs, meaningful-change threshold $\delta$, and fallback.
\item \textbf{Fit before acquisition.} Estimate every parameter using only
  permitted calibration data; do not purchase or reveal holdout features.
\item \textbf{Perturb the feature.} Hold other inputs fixed, sweep the frozen
  support, and compute $D$ from Equation~\ref{eq:viability} on the complete
  prediction pipeline.
\item \textbf{Classify viability.} Record
  \texttt{calibration\_nonviable} if $D=0$ or
  \texttt{calibration\_too\_weak} if $0<D<\delta$; stop or invoke only the
  frozen fallback.
\item \textbf{Acquire after passage.} If $D\geq\delta$, generate the holdout
  feature, lock the mapping, and evaluate it against the strong baseline and
  cheap control.
\end{enumerate}
\textbf{Interpretation:} passage authorizes evaluation; it does not predict
that the feature will improve out of sample.
\end{minipage}}
\caption{The five-step calibration-viability checkpoint. It uses calibration
data only and places a stop decision before costly holdout-feature acquisition.}
\label{fig:checkpoint}
\end{figure*}

The checkpoint tests a necessary condition for interpretation, not predictive
accuracy. An active mapping may still fail out of sample, as the signed repair
did here. The gate prevents an uninterpretable evaluation; it does not promise
a favorable one.

The cheap baseline is part of the checkpoint, not a decorative robustness
test. Headline count should be specified alongside LLM importance before
scoring because it reveals whether semantic inference adds anything beyond
packet volume. Passing the mechanical gate authorizes an evaluation; it does
not exempt the expensive feature from earning its place against that nearly
free control.

Applied prospectively to this study, all four mappings would have received
\texttt{calibration\_nonviable}. The protocol would have stopped after the 2022
fit, preserved the negative calibration result, and avoided the \$16.17
full-history phase. The amount is modest; its value here is as evidence that a
free gate would have prevented a real paid acquisition.

\FloatBarrier
\section{Limitations}\label{sec:limitations}

The two most direct interpretation threats are temporal. The prompt asked
about importance over one to five trading days, while the target was next-day
risk after news had been collected through the close; prices may already have
absorbed that information. Exact visible dates may also have triggered
memorized outcomes. Premarket, intraday, or date-masked designs require new
scores and a new plan.

The evidence covers one LLM family, one horizon, SPY, and closely related QQQ,
not individual stocks or independent replication. Only one linear mapping was
prespecified.

Calibration used 250 rows from 2022 without comparable earlier v4 years, so a
new prospective window might not reproduce the boundary result. Large moves
were rare: 40 SPY and 46 QQQ cases. No ex-ante power analysis was prespecified;
zero slopes gave the prespecified comparison zero ability to detect a score
effect. Verification was same-author, not outside replication.

There was one stochastic generation per main-sample day. Order robustness was
tested only in the 27-date prompt bakeoff, not across the full history; no
repeated-run reliability, financially literate human agreement, or explicit
measurement-error model validates the score as a stable instrument.

The cheap baseline audit covers headline volume and four packet mechanics, not
GDELT tone, a scheduled-announcement calendar, or a comprehensive classical
text model. The corpus also depends on a fixed term query, source whitelist,
crawl timestamps, fuzzy deduplication, and a top-25 cap. The positive
headline-volume endpoint and negative LLM results may not transport under a
different news construction.

\section{Conclusion}\label{sec:conclusion}

This study's central result is a preventable sequencing failure. Full-history
LLM scoring preceded the 2022 viability fit; calibration then assigned all
four news weights zero. The 856 later scores therefore could not affect the
evaluation.

The contrasting analyses sharpen that conclusion. A signed repair restored the
feature path but produced no improvement after correction for four comparisons.
A near-zero-cost headline count improved one endpoint after the same type of
familywise correction. That cheap-control contrast---not the score's raw
association with market moves---is the relevant diagnostic.

The methodological prescription is mandatory and inexpensive: freeze the
gate, fit on calibration data, perturb the candidate feature, and require a
meaningful forecast response before acquiring its holdout history. Then test
the surviving feature against both a strong forecast and the cheapest credible
representation. The operational rule is: \textbf{fit $\rightarrow$ perturb
$\rightarrow$ acquire only if $D\geq\delta$}. If the gate fails, report the
calibration result and stop.

\fi

\ifappendixnow
\appendix
\section{Exact Prespecified Forecast Equations}\label{app:models}

For technical reproducibility, the rescaled news score and option-implied
one-day variance are
\begin{align}
z_t&=(\mathrm{importance}_t-50)/25,\\
q_t&=\max\{(IV_t/\sqrt{252})^2,10^{-8}\}.
\end{align}
If $\tau_t$ is the trailing large-move threshold, the raw probability is
\begin{equation}
p_{\mathrm{raw},t}=2[1-\Phi(\tau_t/\sqrt{q_t})],
\end{equation}
clipped to $[10^{-6},1-10^{-6}]$. The baseline and news probabilities are
\begin{align}
p_{B,t}&=\operatorname{logit}^{-1}[\operatorname{logit}(p_{\mathrm{raw},t})+a_B],\\
p_{N,t}&=\operatorname{logit}^{-1}[\operatorname{logit}(p_{\mathrm{raw},t})+a_N+b_Nz_t],
\end{align}
where $b_N\in[0,5]$ and parameters minimize 2022 binary log loss. Continuous
variance forecasts are
\begin{align}
h_{B,t}&=q_t\exp(c_B),\\
h_{N,t}&=q_t\exp(c_N+d_Nz_t),
\end{align}
where $d_N\in[0,5]$ and parameters minimize
$\ell(h_t,r_t)=\tfrac12[\log(h_t)+r_t^2/h_t]$ on 2022 data. This is QLIKE up to
an inconsequential positive scale and terms that do not vary across forecasts
\citep{patton2011volatility}.

\section{Reviewer-Response Diagnostic Details}\label{app:reviewer}

This \$0 analysis was frozen before calculation, labeled post-hoc, and run
once on saved inputs. The four sign-unrestricted fits appear in
Panel B of Table~\ref{tab:main-contrasts}, together with the later, separately
frozen post-hoc repair that applied those exact fits to the OOS outcomes.

The score was not constant: the 250 training values ranged from 4 to 98
(mean 70.50, SD 17.43), while the 856 later values ranged from 5 to 97
(mean 69.41, SD 17.34). In the 16-endpoint Bonferroni sensitivity, only the
SPY and QQQ Spearman intervals excluded zero:
$[0.01227,0.25796]$ and $[0.00772,0.23460]$. All other risk-ratio,
standalone, and GARCH intervals included their null. Full precision, seeds,
the preserved 46/48 first verification, and the frozen tolerance correction
are in
\path{First Model/validation/news_importance_v4_autonomous/}
\path{reviewer_response_diagnostics/}.

The signed repair was frozen later as a distinct four-endpoint analysis and
used 98.75\% per-endpoint intervals for 95\% Bonferroni familywise coverage.
It was not folded into the already completed 16-item family because doing so
would retroactively redefine that analysis. Panel B of
Table~\ref{tab:main-contrasts} reports every repair estimate and both interval
families.

The nine dates without a strict-valid score were \mbox{2023-05-12},
\mbox{2023-06-21}, \mbox{2024-07-02}, \mbox{2024-08-16},
\mbox{2024-10-04}, \mbox{2024-11-11}, \mbox{2025-03-31},
\mbox{2025-08-29}, and \mbox{2026-01-26}. Their median VIX and VXN closes were
15.36 and 19.53, versus
16.74 and 20.84 on retained dates. Median next-day absolute returns were
0.398\% for SPY and 0.838\% for QQQ, versus 0.514\% and 0.689\%. None was a
large-move date. Under sharp outcome-aware score bounds, the full 865-date
signed improvements remained in $[-0.0000326,-0.0000163]$ (SPY binary),
$[-0.008957,-0.006752]$ (SPY variance),
$[-0.000931,-0.000683]$ (QQQ binary), and
$[-0.009974,-0.007940]$ (QQQ variance).

\section{Additional Exploratory Diagnostics}\label{app:additional}

All analyses in this appendix were designed after the prespecified result was
known. They illustrate the case rather than provide independent confirmation.

\begin{table}[t]
\centering
\scriptsize
\caption{Exploratory relationship between importance and next-day absolute
return on 856 later dates. Intervals are the original unadjusted 95\%
intervals. The risk ratio (RR) compares high- with lower-importance days.}
\label{tab:association}
\begin{tabular}{@{}lcc@{}}
\toprule
Asset & Spearman [95\% CI] & RR [95\% CI] \\
\midrule
SPY & $0.1413\ [0.0564,0.2221]$ & $2.2945\ [0.9191,3.9723]$ \\
QQQ & $0.1284\ [0.0492,0.2022]$ & $1.9735\ [1.0306,3.4104]$ \\
\bottomrule
\end{tabular}
\end{table}

Both rank correlations are small and positive, and their ordinary intervals
exclude zero. The ordinary QQQ risk-ratio interval excludes one; the SPY
interval does not. QQQ shares dates and broad shocks with SPY and is not an
independent second experiment.

\subsection{Conditional Rank Association}

We assigned average ranks to ties, centered the ranks, projected both ranked
importance and ranked absolute return on the same ranked controls, and
correlated the residuals. This partial rank correlation is not a causal
mediation estimate.

\begin{table}[t]
\centering
\small
\caption{Exploratory rank correlations on the same 856 later dates.}
\label{tab:partial-rank}
\begin{tabularx}{\columnwidth}{@{}Xrr@{}}
\toprule
Rank statistic & SPY & QQQ \\
\midrule
Importance vs.\ next $|r|$ & 0.1413 & 0.1284 \\
Importance vs.\ IV & 0.3309 & 0.2959 \\
IV vs.\ next $|r|$ & 0.2742 & 0.2639 \\
Importance vs.\ next $|r|$, conditional on packet mechanics & 0.1393 & 0.1269 \\
Importance vs.\ next $|r|$, conditional on IV & 0.0557 & 0.0546 \\
Conditional on packet mechanics and IV & 0.0486 & 0.0521 \\
Conditional on IV and GARCH variance & 0.0557 & 0.0561 \\
\bottomrule
\end{tabularx}
\end{table}

Accounting for option-implied volatility reduced the raw relationship by
60.6\% for SPY and 57.5\% for QQQ, leaving correlations near 0.055. Jointly
controlling for pre-deduplication volume, domain diversity, mean headline
length, and median crawl age barely changed the raw correlations. No
uncertainty interval was predefined for this later conditional analysis.

\subsection{GARCH Forecast-Loss Comparisons}

The addendum fit Student-t GARCH(1,1) on the previous 1,260 returns and refit
every 21 sessions. We compared GARCH alone and GARCH plus VIX/VXN, each with
and without importance. For every asset, endpoint, and baseline, the mapping
was refit on the same 250 usable 2022 score dates. The news weight was again
restricted to $[0,5]$. If its training-loss improvement was no more than
$10^{-12}$, the rule selected the exactly nested zero-weight model. Parameters
were then fixed on all 856 later dates.

\begin{table*}[t]
\centering
\scriptsize
\caption{All eight post-hoc GARCH comparisons on 856 later dates. Improvement
is baseline loss minus the corresponding model's loss after adding importance;
positive values favor importance. Each news slope was fit separately on 2022
under the $[0,5]$ restriction.}
\label{tab:garch}
\begin{tabular}{@{}lllrrr@{}}
\toprule
Asset & Endpoint & Baseline & News slope & Improvement & 95\% interval \\
\midrule
SPY & Binary & GARCH & 0.003728 & $0.00000398$ & $[-0.00002764,0.00003299]$ \\
SPY & Binary & GARCH + VIX & 0.021228 & $-0.00008905$ & $[-0.00053035,0.00029680]$ \\
SPY & Variance & GARCH & 0 & $0$ & $[0,0]$ \\
SPY & Variance & GARCH + VIX & 0 & $0$ & $[0,0]$ \\
QQQ & Binary & GARCH & 0 & $0$ & $[0,0]$ \\
QQQ & Binary & GARCH + VXN & 0 & $0$ & $[0,0]$ \\
QQQ & Variance & GARCH & 0 & $0$ & $[0,0]$ \\
QQQ & Variance & GARCH + VXN & 0 & $0$ & $[0,0]$ \\
\bottomrule
\end{tabular}
\end{table*}

Only the two SPY binary fits used nonzero importance weights, and both OOS
intervals included zero. The other six fits selected zero, producing identical
news and baseline forecasts. No GARCH comparison showed a clear gain.

\subsection{Standalone Forecast, Missing Dates, and Multiplicity}

A frozen \$0 test removed GARCH and VIX/VXN and used importance alone. At
ordinary 95\% coverage, binary loss improved by 0.00440 for SPY
($[0.00094,0.00789]$) and 0.00185 for QQQ ($[0.00054,0.00313]$), while
continuous volatility did not improve. Binary loss nevertheless remained far
worse than GARCH or VIX/VXN (SPY: 0.328 versus 0.178/0.179; QQQ: 0.339 versus
0.191/0.190). The scores weakly ordered quiet and active days but did not make
a competitive forecast.

The nine missing dates were similar to retained dates in headline count,
domain diversity, length, and crawl age, and none was a large-move date. Five
fixed score scenarios left the raw SPY rank correlation between 0.1338 and
0.1469 and QQQ between 0.1182 and 0.1310. Outcome-aware best cases could not
make any signed-repair point estimate positive; exact bounds appear in
Appendix~\ref{app:reviewer}.

For a frozen 16-endpoint sensitivity---two rank correlations, two risk ratios,
four standalone improvements, and eight GARCH comparisons---we used
99.6875\% per-endpoint intervals to obtain 95\% Bonferroni familywise coverage.
Only the two raw rank correlations remained positive. The signed repair and
cheap-volume analysis were later, separately frozen four-endpoint families and
used 98.75\% per-endpoint intervals. We preserve these analysis-time families
rather than retroactively redefine them.

\section{Exact News-Construction Rules}\label{app:news-construction}

\begingroup
\raggedright
The 26 accepted domains were \texttt{reuters.com}, \texttt{apnews.com},
\texttt{wsj.com}, \texttt{cnbc.com}, \texttt{marketwatch.com},
\texttt{barrons.com}, \texttt{businessinsider.com}, \texttt{forbes.com},
\texttt{economist.com}, \texttt{investing.com}, \texttt{foxbusiness.com},
\texttt{nytimes.com}, \texttt{washingtonpost.com}, \texttt{usatoday.com},
\texttt{latimes.com}, \texttt{theguardian.com}, \texttt{bbc.com},
\texttt{bbc.co.uk}, \texttt{cnn.com}, \texttt{nbcnews.com},
\texttt{cbsnews.com}, \texttt{foxnews.com}, \texttt{npr.org},
\texttt{politico.com}, \texttt{thehill.com}, and \texttt{axios.com}.

The case-insensitive whole-word title query was the union of: federal reserve,
FOMC, central bank, interest rate, interest rates, rate cut, rate hike,
inflation, CPI, consumer price index, jobs report, nonfarm payrolls, payrolls,
unemployment, GDP, recession, stimulus, stock market, stocks, equities, wall
street, S\&P 500, Dow Jones, Dow, Nasdaq, stock futures, treasury yields,
treasuries, bond market, bond yields, tariff, tariffs, trade war, economy, oil
prices, crude oil, bear market, and circuit breaker. Ranking, deduplication,
source caps, timestamps, and update handling are described in
Section~\ref{sec:data}; executable constants and tests are in the ancillary
archive.
\par
\endgroup

\section{Detailed Procedural Chronology}\label{app:chronology}

Four prompts were compared on 27 dates, three per year in 2017--2021 and
2023--2026; 2022 was reserved. Each prompt saw two fixed headline orders and
was judged under a predetermined priority rule. The first API attempt produced
no answers because the requested schema used unsupported
\texttt{uniqueItems}. A technical-only retry removed that setting while keeping
the uniqueness check in local validation. It completed 216 requests, of which
193 passed every format rule.

The initial analysis divided complete paired answers by all expected answers,
although a separate rule already permitted up to 2\% missingness. P1 had 53
valid answers out of 54, but this double-counting reported 0.963 coverage and
made every prompt fail. After LLM outputs but before market outcomes were
opened, Analysis Correction 01 divided by valid responses. P1 then had 1.000
paired coverage and alone cleared the separate validity threshold by one
response: 53/54 = 0.9815 versus 0.98. The correction changed STOP to GO, so
prompt selection was not fully finalized before all LLM-output inspection.

P1 next scored 251 eligible 2022 dates. The first pass produced only 235
strict-valid answers (93.6\%), a formal STOP. Its SPY rank correlation was
0.1133 with a 90\% block interval $[0.0047,0.2106]$. After that result was
known, an audit found that all 16 invalid answers, and no valid answer, had hit
the 8,192-token ceiling. The written plan already allowed one retry for
token-truncated output, so exactly those 16 were rerun with a 16,384-token
ceiling. Fifteen recovered; there was no second retry or manual repair. Final
coverage was 250/251, the rank correlation was 0.0980 with interval
$[0.0045,0.1867]$, and the high-versus-lower importance risk ratio was 1.2544
with interval $[0.8163,1.8530]$. The screen passed because the correlation
interval excluded zero, although the retry weakened rather than strengthened
the association.

Before any 2023--2026 v4 LLM output or QQQ/VXN outcome was opened, the project
fixed the P1 prompt, model, schema, 865 dates, 2022-only fitting,
2023-01-03 through 2026-06-30 evaluation, SPY/VIX and QQQ/VXN endpoints,
missing-output and retry rules, bootstrap and classification rules, code
hashes, and an \$18 full-history cap. Hybrid v1.10 had already evaluated SPY
outcomes over part of the same calendar period. Five batches of 173 requests
produced 815 first-pass valid answers. Exactly 44 token-truncated answers were
retried once and 41 recovered, leaving 856/865 strict-valid dates. This phase
cost \$16.1749105; the complete v4 program cost \$25.0817470, or \$25.08 rounded
to cents. Only after score acquisition did the project open the training-only
implied-volatility snapshots, execute the 2022 calibration, and run the v4
evaluation. These amounts are operational records, not claims of economic
importance.

Same-author verification programs independently reconstructed requests,
retries, hashes, costs, date alignment, fitted models, losses, intervals, and
classifications from saved artifacts. The full-history, GARCH,
conditional-rank, and packet-baseline verifiers passed 18, 17, 13, and 13
checks. A reviewer-response verifier first passed 46/48 checks; the two
failures were optimizer locations differing by at most $3.12\times10^{-8}$
while losses agreed within $6.67\times10^{-16}$. That failure and a frozen
numerical-tolerance correction were preserved. These are software
reimplementations, not third-party replication.

\begin{table*}[t]
\centering
\footnotesize
\caption{Decision chronology. The first market result was known before the
mechanical ceiling retry; all later diagnostics were post-hoc. Paid costs are
displayed to four decimal places; their exact total is \$25.0817470.}
\label{tab:chronology}
\begin{tabularx}{\textwidth}{@{}>{\raggedright\arraybackslash}p{2.85cm}>{\raggedright\arraybackslash}X>{\raggedright\arraybackslash}Xr@{}}
\toprule
Stage & Evidence then available & Result & Cost \\
\midrule
Prompt bakeoff & LLM outputs; no market data & Original STOP; corrected P1 selection & \$4.2219 \\
2022 first pass & SPY outcomes & Formal coverage failure, 235/251 & \$4.3746 \\
Token-limit retry & First-pass market result known & 250/251; Spearman fell to 0.0980 & \$0.3103 \\
Full-history inference & No v4 outputs or QQQ/VXN outcomes opened; prior SPY exposure disclosed & 856/865 usable scores & \$16.1749 \\
Prespecified evaluation & Frozen scores and market snapshots & Four zero slopes; identical forecasts & \$0 \\
Post-hoc GARCH & Main null known & Raw association; no resolved loss gain & \$0 \\
Conditional-rank audit & All preceding results known & Association attenuated to about 0.055 & \$0 \\
Reviewer-response audit & Frozen saved artifacts & Negative free-sign slopes; familywise check & \$0 \\
Signed-mapping repair & Prior results and unrestricted 2022 slopes known & Four active mappings; all estimates nonpositive & \$0 \\
Packet/missingness audit & All prior results known; rules frozen before calculation & Cheap volume mixed; missing-score bounds negative & \$0 \\
\addlinespace
\textbf{Total paid v4 program} & \multicolumn{2}{l}{Sum of the exact stage costs} & \textbf{\$25.0817} \\
\bottomrule
\end{tabularx}
\end{table*}

\section{Reproducibility and Evidence Map}\label{app:evidence}

All paths below are relative to the project root. The release manifest records
the SHA-256 digest of each included artifact.

\begingroup
\raggedright
The canonical root is
\path{First Model/validation/news_importance_v4_autonomous/}. Its included
\path{RUN_LOG.md} provides the comprehensive v4 chronology, corrections, and
costs;
\path{full_history_2023_2026/} contains the prespecified result; and the
\path{posthoc_garch_addendum/}, \path{posthoc_standalone_importance/},
\path{posthoc_conditional_rank_diagnostic/},
\path{reviewer_response_diagnostics/},
\path{posthoc_signed_mapping_repair/}, and
\path{posthoc_packet_baseline_and_missingness/} directories contain each later
specification, result, and verifier.
\endgroup

Core frozen hashes are recorded in the corresponding reports and machine
manifests. The private project retains saved provider outputs as audit inputs,
but they are not redistributed. The public archive supports recalculation of
the reported statistics from released derived rows, not end-to-end recreation
of headline collection or stochastic model scoring. A new API run would be a
new experiment.

\section{Data, Code, and Disclosure Statements}\label{app:disclosures}

The private project retains frozen requests, saved provider outputs, parsed
scores, market snapshots, results, code, and verification artifacts. New API
scoring would be a new experiment. The ancillary archive contains the
redistributable plans, code, derived rows, machine results, verification
outputs, and a SHA-256 manifest. It supports statistical recalculation from
derived rows, not full pipeline replay. Restricted headline text, provider
requests and responses, and raw market downloads are excluded and remain
subject to their source terms.
Internal hashes identify recorded files but do not independently timestamp the
plans.

The human author reviewed the decisions, code, results, and sources and takes
responsibility for the manuscript. The LLM evaluated as the research
instrument is described in Section~\ref{sec:data}.

\ifnamedversion
\section{Funding and Competing Interests}

The author funded the API charges and reports no external funding or competing
interests.
\fi

\fi

\bibliography{references}

\appendixnowtrue

\end{document}